\documentclass[usenatbib]{mn2e}
\usepackage{times}

\title[Hierarchical Merging, Ultraluminous and Hyperluminous X--ray
Sources] {Hierarchical Merging, Ultraluminous and Hyperluminous X--ray
Sources}

\author[A.R.~King and W.~Dehnen]
{
        A.R.~King and W.~Dehnen\\
        Theoretical Astrophysics Group,
        Department of Physics \& Astronomy,
        University of Leicester,
        Leicester, LE1~7RH
}

\begin{document}

\maketitle

\begin{abstract}
Various arguments strongly suggest that the population of
ultraluminous X--ray sources (ULXs: apparent X--ray luminosity $>$
Eddington limit for $10M_\odot$ $\simeq 10^{39}$~erg~s$^{-1}$) in nearby
galaxies are mostly stellar--mass X--ray binaries in unusual evolutionary
stages. However there are indications that the very brightest systems
may be difficult to explain this way.  Accordingly we consider the
class of hyperluminous X--ray sources (HLXs) (i.e. those with apparent
bolometric luminosities $\ga 10^{41}$~erg~s$^{-1}$). Because this class is
small (currently only the M82 object is a secure member) we do not
need to invoke a new formation mechanism for its black holes. We
explore instead the idea that HLXs may be the nuclei of satellite
galaxies captured during hierarchical merging. The observed
correlation between AGN and tidal interactions implies that HLX
activity would switch on during passage through the host galaxy, close
to pericentre. This suggests that HLXs should appear near the host
galaxy, be associated with star formation, and thus possibly with
ULXs.

\end{abstract}


\section{Introduction}
In recent years observations of external galaxies have revealed a
significant population of non--nuclear X--ray sources whose apparent
X--ray luminosities exceed the Eddington limit for a $10M_\odot$ black
hole, i.e.\ apparent $L_X \ga 10^{39}$\,erg\,s$^{-1}$. Many of these
ultraluminous X--ray sources (ULXs) are seen to vary significantly,
ruling out the possibility of superpositions of fainter sources.
Several authors \citep[see][for a recent review]{ColbertMiller} have
suggested that ULXs might contain intermediate--mass black holes
(IMBHs), with masses higher than those resulting from current stellar
evolution (so that the apparent luminosity becomes sub--Eddington) but
still below the supermassive values found in active galactic nuclei.

Such a large population of IMBHs clearly requires a new formation
mechanism distinct from the familiar ones of normal stellar evolution
and supermassive black hole growth in galactic centres, and there have
been several suggestions for such mechanisms. For example primordial
(Population III) stars devoid of metals may have formed such objects
in the early history of the Galaxy \citep{MadauRees}. Alternatively,
IMBHs might be born in dense star clusters, either as the result of
mergers of stellar--mass black holes \citep{MillerHamilton}, or of
stellar mergers on a timescale too short for nuclear evolution
\citep{Gurkanetal, PortegiesZwartetal, Hopmanetal}. However there are
objections to all of these suggestions. Primordial IMBHs need to find
themselves non--primordial reservoirs (probably stars) to accrete
from. Black--hole mergers are subject to gravitational radiation
recoil which probably ejects them from the cluster
\citep{Merritt:recoil, Madau:recoil} before the mass can grow
significantly, and stellar mergers may provoke significant mass loss
which limits mass growth.

Here we take a different view. In the next Section we summarize the
observational arguments why the majority of ULXs are probably
stellar--mass X--ray binaries rather than accreting IMBHs. However
there remain a few very bright sources where there may be
significantly higher masses. Since this group is small, there is no
compelling reason to invoke a new formation mechanism for its
members. Accordingly we consider the possibility that these black
holes are formed in the second way already familiar to us, i.e. in the
centres of galaxies.

\section{Stellar--mass or intermediate--mass black holes?}

Here we summarize the evidence that ULXs are in the main stellar--mass
X--ray binaries rather than IMBH. The X--ray luminosity function of
nearby galaxies, normalized by star--formation rate, shows no break at
$L_X \sim 10^{39}$~erg~s$^{-1}$ \citep{Grimmetal}. This strongly
suggests that most ULXs are simply X--ray binaries in some unusual
shortlived phase. Some stellar--mass X--ray binaries, e.g.\
GRS~1915+105 \citep[see][b]{Doneetal,Bellonietala} are indeed observed
to radiate at apparent luminosities above their Eddington limits. The
X--ray spectral temperatures $\sim 1-2$~keV are more easily compatible
with stellar--mass than intermediate--mass black holes. The very
strong association between ULXs and induced star formation epitomised
by the Antennae \citep[][ 2004]{Fabbianoetal3} and Cartwheel galaxies
\citep{Gaoetal} is not easy to explain in IMBH models. In particular
\citep{King04} the spreading ring of star formation seen in the
Cartwheel requires at least 300 `dead' IMBHs to have formed inside it,
demanding $\ga 10^8M_\odot/\eta$ in clusters if IMBHs formed in them,
where $\eta$ is the efficiency for an IMBH to find a companion.  These
number increase considerably if either the distribution of IMBH or
companion stars grow exponentially within the ring, and are in any
case underestimates unless we are observing at a special epoch where
the ring has passed the last few IMBH and there are no more at larger
radii. (The argument in \citep{King04} that ULXs containing IMBH must
be largely transient, thus raising the numbers by a further factor
(duty cycle)$^{-1}\ga 10$, does not hold if the companions are
sufficiently massive.) Finally, many ULXs are observed just outside
star clusters \citep{Kaaretetal04} compatible with supernova kicks for
stellar--mass black holes, but not with IMBH formation in the cluster.

In light of some of this evidence \cite{Kingetal} suggested instead
that most ULXs are high--mass X--ray binaries in which the donor star
is initially more massive than the black hole and fills its Roche
lobe. As a result mass transfer is on a short (thermal) timescale, and
thus super--Eddington. Population synthesis studies by
\cite{Podsiadlowski} show that the subsequent mass transfer on the
star's nuclear time is also high enough to explain many ULXs, and
indeed produces systems with a longer lifetime which may dominate the
ULX population numerically. In this phase the black hole mass can grow
by up to a factor 2, and the hole ultimately accretes material of low
hydrogen abundance, both effects increasing the value of the Eddington
limit (see below). \cite{King02} pointed out the existence of a second
group of systems where super--Eddington accretion rates occur, namely,
the unstable accretion discs driving long--lasting outbursts of soft
X--ray transients. These would necessarily provide the stellar--mass
ULXs in elliptical galaxies, where no high--mass X--ray binaries
remain.

Given super--Eddington accretion rates in stellar--mass systems, there
are three possible avenues for explaining the high apparent X--ray
luminosities of ULXs. The high accretion rates may cause disc warping
\citep{Pringle} or other forms of scattering of the emitted X--rays.
The resulting mildly anisotropic emission can then appear
super--Eddington on some lines of sight
\citep{Kingetal}. Alternatively, accretion may genuinely be
super--Eddington, either because the disc atmosphere is partly
magnetically structured \citep{Begelman} or because it becomes
super--Eddington at large disc radii \citep{ShakuraSunyaev}. In all
three cases the $L_X$ inferred from observation can be $\sim10$ times
larger than the classical isotropic Eddington limit. The twin effects
of black hole mass growth and accretion of hydrogen--depleted material
referred to above raise the formal Eddington limit to a value
\begin{equation}
L_{\rm E} = 4.4\times 10^{39}M_{20}~{\rm erg~s}^{-1}
\end{equation}
where $M_{20}$ is the black hole mass in units of $20M_\odot$.  Hence
stellar--mass models can explain ULX bolometric luminosities up to a
value
\begin{equation}
\sim 4 \times 10^{40}~{\rm erg~s}^{-1} 
\end{equation}
without real difficulty.

\section{ULXs and HLXs}

The observational considerations summarized above strongly suggest
that {\it most} ULXs are probably stellar--mass X--ray binaries in
unusual phases. However \cite{Kingetal} emphasize that these kinds of
population argument cannot rule out the possibility that {\it a few}
systems might contain IMBHs. In particular, the ULX in M82
\citep{Matsumotoetal,Kaaretetal} is very bright ($L_X \sim 10^{41}~{\rm
erg~s}^{-1}$) for stellar--mass models. There are also a number of
interesting, although not clinching, arguments that suggest a more
exotic origin for the very brightest ULXs \citep[for a review
see][]{ColbertMiller}.  To avoid confusion we therefore follow
\cite{Gaoetal} and consider {\it hyperluminous X--ray sources} (HLXs)
as those with apparent $L_X \ga 10^{41}$~erg~s$^{-1}$. To date, the
only secure member of the HLX class is the M82 object. Although the
brightest of the Cartwheel sources formally exceeds the defining
limit, there is as yet no demonstration (e.g. of variability) which
would firmly rule out a superposition of fainter sources.  We retain
the designation ULX for systems with lower apparent $L_X$. We note
that the luminosity of the HLX class is comparable with that of the
intrinsically faintest active galactic nuclei (AGN).

Accepting that most if not all ULXs are stellar--mass binaries leaves
us to explain the HLXs. The incidence of this class is at most one per
several galaxies. Accordingly, finding a model for it is a much more
tractable task than inventing a new model for the entire ULX
class. Indeed Occam's razor suggests that we should not look for new
ways of making and feeding black holes, but instead consider ways of
using the black holes we already know of.

\section{HLXs and Hierarchical Merging}

Black hole formation and feeding is well established in two contexts:
\begin{itemize}
        \item[(i)]  \label{A} stellar--mass binaries, and
        \item[(ii)] \label{B} the centres of galaxies.
\end{itemize}
By hypothesis we are abandoning attempts to explain HLXs (as opposed
to ULXs) in terms of stellar--mass black holes, so we can only use
(ii). The obvious possibility to consider is that the hierarchical
merger picture of structure formation \citep{WhiteRees} predicts that
in the present Universe, large galaxies have captured between 10 --
100 dwarf satellite galaxies. If a sufficient fraction of these
satellites have retained their central black holes, and the attendant
structure such as the dense star clusters probably implicated in
X--ray activity, we may expect some of them to become active and
appear as HLXs from time to time. In the simplest picture of
sub--Eddington accretion and isotropic X--ray emission, any black hole
of mass $M_{\rm BH} \ga 170M_\odot$ (for accretion of hydrogen--rich
matter) is a candidate for explaining HLXs, provided it can be fed at
a rate $\sim 10^{-5}M_\odot$~yr$^{-1}$.

This picture offers a straightforward explanation for the required
black hole mass. The observed $M_{\rm BH} - \sigma$ relation
\citep{FerrareseMerritt,Gebhardtetal,MerrittFerrarese} suggests that
even a dwarf spheroidal with velocity dispersion $\sim 20$~km~s$^{-1}$
would have $M_{\rm BH} \sim 10^4M_\odot$. We note that explanations of
the $M_{\rm BH} - \sigma$ relation in terms of self--limited black
hole growth during galaxy formation \citep{SilkRees,King03}
work independently of the value of $\sigma$.

The mechanism by which accretion switches on in AGN is not yet well
understood, and the same must hold for the satellite nuclei we
consider \citep[for discussions see e.g.][and references therein]{
Taniguchi1999, CavaliereVittorini2000, KewleyDopita2003, Corbin2000,
ViraniEtal2000, DeRobertisEtal1998}. There has been considerable
discussion of the idea that the activity results from the capture and
disruption of a small satellite galaxy. In this paper we investigate
what consequences this idea might have for HLXs.

The satellite galaxies must have very eccentric orbits about the host
galaxy. Only those approaching close to the centre of the host will
feel strong tides. Deduced HLX accretion rates consume a star in only
$10^6 - 10^7$~yr, short compared with the orbital timescale near
pericentre. So any activity must necessarily occur only when the
satellite is very close to the centre of the host galaxy. Moreover,
the passage of the satellite through the host must trigger star
formation \citep{MihosHernquist}, as spectacularly observed in the
Cartwheel \citep{Gaoetal}. This kind of activity leads in turn to ULX
formation on a timescale $\sim 10^7$~yr \citep[cf][]{King04}. Thus our
picture naturally predicts that HLXs occur near their host galaxy, and
may be accompanied by starburst phenomena such as ULXs.

\subsection{How close to the host do we expect HLXs?}
In order to answer this question, we need to know how the satellite
nucleus is activated. Since the mechanisms responsible for feeding AGN
are not well understood, we cannot make any precise statements.
However, since large impact distances are much more likely, it is
clear that a mechanism which works with smaller tides is preferred.

If the satellite has any remaining gas, it may be channeled to the
nuclear region either directly by the tides due to the host galaxy or
by a stellar bar in the satellite the formation of which was triggered
by the flyby. However, dwarf galaxies generally host little if any
gas.  Alternatively, the tidal forces may push several satellite stars
onto radial orbits, so that they will either feed the BH directly or
stir the nucleus sufficiently to activate a dormant accretion disk.
Finally, the BH may feed directly on the gas or stars of the host
galaxy.

We now estimate the impact distance $R$ required for the tidal forces
to generate significant perturbation to the central region of the
satellite. The tidal force generated at projected distance $x$ from
the centre of the satellite which is at impact distance $R$ from the
centre of the host is $F_t\approx xv^2_h/R^2$, where $v_h$ is the
circular speed of the host. For an encounter with velocity $V$, this
force acts during a time $\Delta t\sim 2R/V$ and, according to the
impulse approximation, generates a velocity change $\Delta v\sim F_t
\Delta t=2xv_h^2/RV$.  This velocity change affects only stars for
which $\Delta t$ is shorter than their dynamical time
$t_{\mathrm{dyn}}$, for otherwise the tidal force changes are
adiabatic. If the satellite dwarf galaxy has a central (1D) velocity
dispersion $\sigma_s$ and core radius $r_s$, then the dynamical time
in its centre is $t_{\mathrm{dyn}}\sim 2r_s/\sigma_s$ and the tidal
shock is impulsive for all stars if
\begin{equation} \label{eq:dist:dt}
  R \la \frac{r_s V}{\sigma_s}.
\end{equation}
We may now estimate the relative change of the kinetic energy of the
satellite core (assuming impulsive shock for all stars)
\begin{equation}
  \frac{\langle\Delta E\rangle}{E}
  = \frac{\langle v\Delta v + \frac{1}{2}(\Delta v)^2\rangle}
           {\frac{3}{2}M\sigma_s^2}
  \simeq \frac{4}{15}\frac{r_s^2v_h^4}{R^2V^2\sigma_s^2}
\end{equation}
where we have used $\langle x^2\rangle=r_s^2/5$ for a near homogeneous
density core. The relative change $\langle(\Delta E)^2\rangle/E^2$ is
even a factor 2 larger. If we now assume that the encounter occurs at
a distance as given by equation~(\ref{eq:dist:dt}), we find
\begin{equation} \label{eq:dE}
  \frac{\langle(\Delta E)^2\rangle}{E^2} \ga
  \frac{8}{15} \left(\frac{v_h}{V}\right)^4.
\end{equation}
For typical values of $v_h\sim 200\,$km\,s$^{-1}$ and $V\sim
500\,$km\,s$^{-1}$, the right-hand side of equation~(\ref{eq:dE})
amounts to only 1.4\%. Thus, the condition that the shock is impulsive
for all stars does not necessarily mean that it is strong. For the
shock to be strong ($\langle(\Delta E)^2\rangle/E^2\ga10\%$), we
require
\begin{equation} \label{eq:dist:dE}
  R \la 2 \frac{r_sv_h^2}{V\sigma_s}.
\end{equation}
If we assume $r_s\sim 100\,$pc, $\sigma_s\sim20\,$km\,s$^{-1}$ and
values for $v_h$ and $V$ as used above, we find $R\la1\,$kpc for the
tidal forces to significantly perturb the inner regions of the
satellite galaxy. 

Clearly, at larger distances from the satellite centre, the tidal
perturbations are stronger and likely to be disruptive. We should note
that that close to the host any direct observations of the stellar
body of the satellite are difficult if not impossible.

The one clear member of the HLX class, namely the M82 source, is about
200~pc from the nucleus, in agreement with our rough estimate
above. There are two other objects which come into consideration as
HLXs. The ULX in NGC2276 \citep{DavisMushotzky} has a 0.5--2.0~keV
luminosity of $3.2\times 10^{40}$~erg~s$^{-1}$, extrapolated to a
0.5--10~keV luminosity of $1.1\times 10^{41}$~erg~s$^{-1}$. The source
is seen to vary between $2.2\times 10^{40}$~erg~s$^{-1}$ and
$4.4\times 10^{40}$~erg~s$^{-1}$ in the 0.5--2.0~keV band, suggesting
a varying source of at least $5.5\times 10^{40}$~erg~s$^{-1}$. This
source is in the outer disc of the galaxy. The colliding galaxy
NGC7714 has a source with luminosity $7\times 10^{40}$~erg~s$^{-1}$ in
XMM data \citep{SoriaMotch} extrapolated to a bolometric luminosity of
$\sim 1.5\times 10^{41}$~erg~s$^{-1}$. The source is observed to vary
by a factor 2, suggesting a luminosity of at least $\sim 7.5\times
10^{40}$~erg~s$^{-1}$. This source is at the junction of the tidal
tail from the colliding galaxy NGC7715 with the collisional star
formation ring of NGC7714, and is about 4kpc from the nucleus of the
latter. It is not itself in a region of star formation. Both objects
are formally below our (slightly arbitrary) luminosity limit for HLXs.

\subsection{The estimated frequency of HLXs}

We may try to estimate the frequency of HLXs per host galaxy from the
frequency of satellite galaxies close to the centre of the host.
Simulations of hierarchical structure formation consider dark matter
only, and usually find the number density of sub--haloes to be
constant for the inner regions of a host halo
\citep[e.g.][]{DBS04}. From the high--resolution galaxy halo
simulations of these authors, we estimate $\sim10^{-6}\,$kpc$^{-3}$
subhaloes with masses $\sim10^{-4}$ that of their host. If a distance
of $\sim1\,$kpc is necessary to trigger an HLX, we get an HLX
frequency of $\sim10^{-6}-10^{-5}$ per host. This estimate is rather
crude because the simulations do not include the galaxies (stars and
gas) and their influence on the dynamics of sub--structure.

One might instead try to use observational constraints on the number
of satellite galaxies in the close vicinity of their hosts. However,
for distances as close as required here observations are incomplete,
and there is very little information in the literature. If we assume
that structure is scale--free as in CDM simulations,
\citep[e.g.,][]{CYE97,LM03} extrapolating from the number density of
galaxies in clusters suggests a number of satellites per galaxy some
100 to 1000 times larger.

We thus conclude that the frequency of HLX per host galaxy lies
somewhere between $\sim10^{-6}$ and $10^{-2}$. Our limited knowledge
of substructuring on kpc scales prevents a more accurate
estimate. Observations of HLXs, if indeed caused by IMBHs at the
centres of satellite galaxies, may prove useful for understanding
both the formation of satellite galaxies and the triggering of AGN.

\section{Discussion}

Our explanation of the brightest non--nuclear X--ray sources has
followed two stages; first, the recognition that the ULX population
($10^{39}$~erg~s$^{-1} \la$ apparent $L_X \la 10^{41}$~erg~s$^{-1}$)
is mainly a collection of stellar--mass binaries in unusual states;
and second, the identification of the HLX class (apparent $L_X \ga
10^{41}$~erg~s$^{-1}$) as a much smaller group of possible
higher--mass systems. The resulting low incidence of HLXs per galaxy
led us to explore the idea that they could represent the nuclei of
some of the satellite galaxies predicted by hierarchical merging. The
observed correlation between AGN and tidal interactions suggests that
HLX activity is switched on by passage through the host galaxy, close
to pericentre. This suggests that HLXs should be associated with star
formation and thus possibly with ULXs. Further tests of our idea
exploring the connection with the merger history of galaxies may have
to wait for the accumulation of a larger sample of HLXs.

IMBH models of the entire ULX class inevitably have to postulate a new
mode of black hole formation and a new method of feeding the hole. By
separating the small HLX class from the majority of ULXs we can
instead confine ourselves to formation and feeding processes which are
well established from observations of X--ray binaries and AGN. In
addition, the proposed link to hierarchical merging suggests that HLX
may have something to tell us about galaxy formation.

\section*{Acknowledgments}

Theoretical astrophysics research at Leicester is supported by a PPARC
rolling grant. We thank several members of this group, as well as
Philipp Podsiadlowski, Vicky Kalogera and Cole Miller, for helpful
discussions and the anonymous referee for useful comments. This paper
was partly written during a visit to the Astronomical Institute,
University of Amsterdam, and ARK thanks members of that institute for
hospitality and discussions. ARK gratefully acknowledges a Royal
Society Wolfson Research Merit Award.

\end{document}